\documentstyle[12pt]{article}
\begin{document}
\vspace*{-2mm}
\thispagestyle{empty}
\noindent
\hbox to \hsize{
\hskip.5in \raise.1in\hbox{\bf University of Wisconsin - Madison}
\hfill$\vcenter{\hbox{\bf MAD/PH/849}
                \hbox{\bf UCD-94-34}
                \hbox{\bf RAL-94-099}
                \hbox{\mbox{}}}$}
\mbox{}
\hfill September 1994   \\
\vspace{0.5cm}
\begin{center}
  \begin{Large}
  \begin{bf}
Further ways to distinguish single-lepton top quark signals from
background at the Tevatron
  \end{bf}
  \end{Large}
  \vspace{0.8cm}

V. Barger and E. Mirkes\\[2mm]
{\it Physics Department, University of Wisconsin, Madison, WI 53706, USA}\\
\vspace{0.3cm}
J. Ohnemus\\[2mm]
{\it Physics Department, University of California, Davis, CA 95616, USA}\\
\vspace{0.3cm}
R.J.N.~Phillips\\[2mm]
{\it Rutherford Appleton Laboratory, Chilton, Didcot, Oxon OX11 0QX, UK}\\[2cm]
{\bf Abstract}
\end{center}
\begin{quotation}
\noindent
Further kinematical variables are suggested, in which to compare the
putative leptonic $W$ plus 4-jet top quark signal with the QCD
background. We show that the lepton rapidity asymmetry,  the
$p_T^{}$-ranking of the  tagged $b$-jet, the double-tag probability,
the reconstructed  $t+\bar t$ invariant mass, and the  lepton energy
in the parent top quark rest-frame, all display interesting
differences between signal and background.
\end{quotation}
\newpage

Evidence for top quark production in $p\bar p$ collisions at the
Fermilab Tevatron collider was recently presented by the CDF
collaboration~\cite{cdf}, corresponding to $t\bar t$ production with
$t\to bW^+$ and $\bar t\to \bar bW^-$; the D0 collaboration has also
presented a top quark candidate event~\cite{d0}.  In the present work
we address the $b$-tagged single-lepton signal~\cite{HAN},
where one  of the $W$
bosons decays leptonically $W\to \ell\nu$ and the other decays
hadronically to two jets $W\to jj$,  while at least one of the four
final jets is tagged as a probable $b$-jet by vertex or
secondary-lepton criteria.  The most serious background to this signal
comes from QCD production of $W+4$ jets;  the CDF event criteria were
designed to suppress this background, but nevertheless it  will be
desirable to make further cross-checks to establish as firmly as
possible the credentials of the putative top quark signal in present
and future data. The CDF collaboration has already presented several
such supplementary checks, notably against the various jet transverse
energy $E_T $ distributions and correlations and  against the summed-$
E_T(j) $  distribution (first advocated in Ref.~\cite{bbp}).

In the present Letter we draw attention to five further quantities,
where the behaviour of top quark signal and background are
significantly different, and illustrate these differences by specific
calculations. The quantities proposed here are:

\begin{enumerate}
\item[(a)] The forward-backward asymmetry in the charged-lepton
rapidity distribution,
\begin{equation}
A(y_{\ell})= \pm [d\sigma /dy(y_{\ell}) - d\sigma /dy(-y_{\ell})]
                /[d\sigma /dy(y_{\ell}) + d\sigma /dy(-y_{\ell})] \>,
\end{equation}
where the $\pm$ sign is equal to the lepton charge and $y > 0$ is the
hemisphere in the $p$ beam direction.

\item[(b)] The ranking-order (1,2,3,4) of the $b$-tagged jets (where
the jets are $p_T^{}$-ordered and jet 1 has highest $p_T^{}$).

\item[(c)] The probability that a tagged event has two tagged jets.

\item[(d)] The invariant mass $m(t\bar t\,)$ of the two top quarks,
extracted by explicitly reconstructing the event from the lepton and
jet momenta and the missing-$E_T$ vector.

\item[(e)] The lepton energy $E_\ell$ in the reconstructed rest-frame
of the semileptonically decaying top quark.
\end{enumerate}

In the following, we shall discuss the qualitative features that cause
signal and background to differ, and give quantitative examples. But
first we describe our methods of calculation.

For our illustrations, we make parton-level Monte Carlo calculations.
We generally calculate $t\bar t$ production at lowest order
($\alpha_s^2$),  with full spin correlations~\cite{spincor} in the
subsequent $W$ boson  decays, using the MRS set D0 parton
distributions~\cite{partons} at  scale $Q=m_t$, taking
$m_t=174$~GeV~\cite{cdf} and normalizing to the
next-to-leading-order total $t\bar
t$ cross section~\cite{laenen}.   However, for the lepton asymmetry,
which vanishes at lowest order, we calculate $2\to3$ parton
subprocesses at order $\alpha_s^3$~\cite{ellsex}, using the truncated
shower approximation~\cite{tsa}. We calculate $W+4$-jet production at
leading order with the VECBOS  program~\cite{vecbos}, supplemented by
the programs developed in Ref.~\cite{bmps},  at scale $Q =\,
<\!p_T^{}(j)\!>$   (the mean jet transverse momentum), with the same
parton  distributions and 4 quark flavors. We set out to impose
approximately the same acceptance cuts as in the CDF single-lepton top
quark search~\cite{cdf}. We regard the final four partons as jets, and
require the transverse momenta $p_T^{}(j)$, pseudorapidities
$\eta(j)$, and separations  $\Delta R(jj)$  of the first three jets to
satisfy
\begin{equation}
p_T^{}(j) > 20\, {\rm GeV},\quad |\eta(j)| < 2.0,
\quad \Delta R(jj) > 0.7,\quad (j=1,2,3),
\end{equation}
where $\eta=\ln(\tan\theta /2)$ and $(\Delta R)^2=(\Delta\eta)^2 +
(\Delta\phi)^2$; $\theta$ and $\phi$ are the usual polar and azimuthal
angles measured with  respect to the antiproton beam direction. These
conditions are relaxed for the fourth jet:
\begin{equation}
p_T^{}(4) > 13\, {\rm GeV},\quad |\eta(4)| < 2.4,\quad \Delta R(j4) > 0.7.
\end{equation}
In the neighborhood of these cuts, studies of jet fragmentation  and
instrumental effects have shown that the initiating parton $p_T$
exceeds the measured jet $E_T$ by about 5~GeV in the
CDF  apparatus~\cite{cdf};
thus the parton cuts above  are intended to approximate
the jet cuts  $E_T(j=1,2,3) > 15$ GeV  and $E_T(4) > 8$ GeV in the CDF
top quark analysis~\cite{cdf}.  For leptons we require
\begin{equation}
p_T^{}(\ell) > 20\, {\rm GeV},\quad |\eta(l)| < 1.0,\quad
    \Delta R(\ell j) > 0.4,\quad (j=1,2,3,4),
\end{equation}
where the lepton-jet separation approximates the lepton isolation
criterion. We apply realistic gaussian smearing factors~\cite{smear}
to  all parton and lepton momenta, to represent measurement errors,
with CDF resolution values~\cite{cdfw}. The missing-transverse energy
vector $E\llap /_T$ is defined to be the negative sum of all the
lepton and parton transverse energy vectors, and must satisfy
\begin{equation}
E\llap /_T > 20\, {\rm GeV}.
\end{equation}
With these cuts, we calculate the total lepton-plus-four-jet signal
from $t\bar t$ production to be 0.50~pb, to be compared with  0.062~pb
from $Wbbjj$ alone and 2.3~pb from all $Wjjjj$ channels.   (If we
choose scale $Q^2 =\, <\!p_T^{}(j)\!>^2 + M_W^2$ instead, these
background numbers become 0.029~pb and 1.1~pb, respectively.) The
signal/background ratio can then be improved by $b$-tagging.

In the CDF experiment, the efficiency for tagging one or more $b$-jets
in a $t\bar t$ event is about 0.33 (combining vertex and lepton
tagging approaches), corresponding to a probability $\epsilon_b \simeq
0.18$ per $b$-jet; the probability of a fake $b$-tag appears to be
$\epsilon_q \simeq \epsilon_g \simeq 0.01$ per light-quark or gluon
jet.  We assume a probability $\epsilon_c \simeq 0.05$  for a bogus
$c$-jet tag.  Tagging efficiencies are taken to be  approximately
independent of $p_T$ and $\eta$. The cross  section for each final
configuration is multiplied by the corresponding probability that at
least one of the jets is tagged; {\it e.g.}, the tagged cross sections
for $Wgggg$, $Wc\bar cgg$, $Wb\bar b qq'$ production contain
tag-probability  factors 0.04, 0.12, 0.34, respectively.   With these
factors, Ref.~\cite{bmps} shows  that the $b$-tagged $W+4$-jet
background comes mainly from $W+4q/g$ final states, {\it i.e.}, from
mistagging light-quark or gluon jets; next, in descending order of
importance, come $W+2b+2q/g$, $W+c+3q/g$, and $W+2c+2q/g$ final
states.  The tagged $W+4$-jet background is therefore dominated by the
$Wjjjj$ and $Wbbjj$ configurations,  which can be calculated from the
VECBOS program~\cite{vecbos} alone. With these tag-factors, the net
tagged $t\bar t$ single-lepton signal becomes 0.13~pb, to be compared
with 0.055~pb (for scale $Q =\,<\!p_T^{}(j)\!>$)
from the combined $Wjjjj$ channels.

We now discuss and illustrate the kinematical distributions (a) -- (e)
described above, for the single-lepton $t\bar t$ signal and the $Wjjjj$
background at the Tevatron.

\subsubsection*{(a) Lepton asymmetry}
For $Wjjjj$ production, two different effects contribute (in opposite
directions) to a forward/backward lepton asymmetry. The preponderance
of valence $u$-quarks over $d$-quarks in  the proton causes $W^+ \,
(W^-)$ bosons to be produced preferentially in the  $p\, (\bar p)$ beam
hemisphere.  The $V-A$ couplings in the subprocess $ud \to W \to
\ell\nu$ produce $\ell^+ (\ell^-)$ preferentially  along the $\bar p\,
(p)$ axis in the $W$ boson rest-frame. At the Tevatron energy the
former effect prevails and the background  $\ell^+ (\ell^-)$ is
predicted to be produced preferentially in  the $p \,(\bar p)$ beam
hemisphere in the lab frame; see Fig.~1. On the other hand, the $p\bar
p \to t\bar t$ tree-level mechanisms produce unpolarized top quarks,
with no forward/backward asymmetry at lowest order;
hence there is no lepton asymmetry here at order $\alpha_s^2$.  At
order $\alpha_s^3$, however, the subprocess $q\bar q \to t\bar
tg$ (which dominates over other $2\to 3$ subprocesses
at the Tevatron energy for $m_t = 174$~GeV)
produces $t$-quarks preferentially in the  direction of the incoming
antiquark $\bar q$, due to interference between gluons emitted from
initial and final quarks~\cite{halzen}; this leads to a small lepton
asymmetry of {\em opposite} sign to the $Wjjjj$ case.  Note that the
$Wjjjj$ effect arises from the isospin asymmetry  within the valence
quarks in $p$ and $\bar p$, whereas the $t\bar t$  effect comes from
the difference between  the triplet and antitriplet color
representations of valence quarks and antiquarks.  The  asymmetries
and the underlying lepton rapidity distributions are shown  in
Figs.~1(a) and 1(b), respectively; both these quantities differ
significantly between signal and background.  If we define the forward
lepton hemisphere as $y(\ell) > 0$ $(y(\ell) < 0)$ for $\ell^+$
$(\ell^-)$, the integrated forward/backward event ratios corresponding
to the asymmetries in Fig.~1 are
\begin{eqnarray*}
&t\bar t: & F/B = 0.95 \>, \\[2mm]
&Wjjjj  : & F/B = 1.25 \>.
\end{eqnarray*}
Good statistics are needed for discrimination here. The present 7
selected top quark candidate events presented by CDF~\cite{cdf} have
ratio $F/B = 3/4$.

\subsubsection*{(b) Rank of tagged jet}

The  $t\bar t \to (b\ell\nu )(bjj)$ signal events contain two genuine
$b$-jets; for $m_t \sim 170$~GeV these jets usually have the highest
$p_T^{}$ and are likely to have ranks 1 and 2 (lowest rank = highest
$p_T^{}$). They are also by far the most likely  jets to be tagged
in a $t\bar t$ event, so
the tagged jet is most likely to have rank 1 or 2 in the signal.  For
the $Wjjjj \to \ell\nu jjjj$ background events, however, we have
argued above that by far the most likely situation is  a bogus tag,
with approximately equal probabilities to have rank 1, 2, 3, or 4.  To
illustrate these expectations, we have calculated  the following
probabilities $P(r)$ that the jet of rank $r$ is tagged:
\begin{eqnarray*}
&t\bar t: & P(1) = 0.34, \quad P(2) = 0.33,
      \quad P(3) = 0.27, \quad P(4) = 0.18, \\[2mm]
&Wjjjj  : & P(1) = 0.27, \quad P(2) = 0.26,
      \quad P(3) = 0.26, \quad P(4) = 0.24.
\end{eqnarray*}
These tagging probabilities have different profiles in the $t\bar t$
and $Wjjjj$ cases; in particular, the ratios of  probabilities
$P(1)/P(4)$  differ significantly:
\begin{eqnarray*}
&t\bar t: & P(1)/P(4)\, = 1.9, \\[2mm]
&Wjjjj  : & P(1)/P(4)\, = 1.1.
\end{eqnarray*}
The seven selected CDF top quark candidate events~\cite{cdf} have  3
cases with rank 1 and just one case with rank 4, giving a first
estimate $P(1)/P(4)=3$.

\subsubsection*{(c) Double-tag probability}

The probabilities $P(r)$ above sum to  more than 1, the excess being
due to multiple tagging, mostly  double-tagging. If we denote the
probability that a tagged event  has $n$ tagged jets by $P(\hbox{\rm
n-tag})$, then
$\Sigma_r P(r) = 1 + \Sigma_n (n-1) P(\hbox{\rm n-tag})$.
$P(\hbox{\rm 2-tag})$ is bigger for  $t\bar t$ signal events (which
always contain two $b$-jets) than for $Wjjjj$ background events (which
mostly arise from bogus tags with no $b\bar b$ pairs present). Direct
calculations with the assumed tagging efficiencies give
\begin{eqnarray*}
&t\bar t: & P(\hbox{\rm 2-tag})\, = 0.12, \\[2mm]
&Wjjjj  : & P(\hbox{\rm 2-tag})\, = 0.034.
\end{eqnarray*}
None of the 7 CDF events is double-tagged in this sense.

\subsubsection*{(d) $t\bar t$ invariant mass}

Candidate $t\bar t$ events can be kinematically reconstructed, using
various constraints.  Equating $E\llap /_T$ to the neutrino transverse
momentum, the $W\to\ell\nu$ kinematics are determined within a
two-fold ambiguity; also $W\to jj$  may be identified by the
best-fitting pair of jets. The remaining two  jets (one of which is
tagged) are then identified as $b$-jets; leptonic decay information
may further identify one of them as $b$ or $\bar b$, but in general
they are not distinguished and there are 4 possible ways of pairing
them with $W\to\ell\nu$ (2 solutions) and $W\to jj$ to form the decay
products  of $t$ and $\bar t$.  We select the assignment in which the
invariant masses of the two candidate top quarks agree most closely,
obtaining a unique ``best fit".   For true $t\bar t$ events, this will
usually give the correct assignment.  For $Wjjjj$
background events, the best fit may in fact be unacceptably poor; in
our analysis we have required the $W\to jj$ candidate to have $|m(jj)
- M_W| < 15$~GeV and the two top quark candidate masses to agree
within 50~GeV.  (Note that our reconstruction method  is not identical
to the CDF procedure, which explicitly rescales momenta, constraining
candidate $W$ masses to equal $M_W$ and candidate top masses to equal
each other.)   Background events that survive these reconstruction
criteria will not necessarily agree with  $t\bar t$ expectations in
their various other kinematical distributions.  Figure~2 illustrates
the differences in the invariant  mass $m(t\bar t\,)$ distributions,
which are known to be sensitive to $t\bar t$
production dynamics~\cite{lane};
the $m(t\bar t\,)$ values for the 7 reconstructed CDF
events are shown by arrows.   The $m(t\bar t\,)$ variable is
loosely related to the summed-$E_T(j)$ variable, since both reflect
different aspects of the total CM energy release, but $m(t\bar t\,)$
has the advantage of including the longitudinal and leptonic degrees
of freedom.

\subsubsection*{(e) Lepton spectra}

The lepton energy $E_\ell$ in the $t$-restframe from $t\to b\ell^+\nu$
decay is determined by the $V-A$ character of the weak couplings; in
particular, it differs characteristically from the lepton spectrum in
the $b'\to c\ell^-\bar\nu$ decay of a hypothetical fourth-generation $b'$
quark.  It is {\it a priori} unlikely that either of these spectra
will be accurately faked by reconstructed background $Wjjjj$ events.
Figure~3 illustrates the differences; the $E_\ell$ values for the 7
reconstructed CDF events are shown  by arrows.  In principle $E_\ell <
{1\over 2}m_t$ from kinematics; the small tail in the signal at
high-$E_\ell$ in Fig.~3 is due to momentum smearing and reconstruction
errors (primarily from errors in reconstructing the longitudinal
momentum of the $W$ boson).  The fourth-generation $b'\bar b'$ curve
is shown simply for comparison; this scenario does not offer an
independent explanation of the CDF events because of the low
probability of tagging $b' \to c W$ decays.

We conclude that all five of the kinematical quantities (a) -- (e)
studied here show significant differences between the single-lepton
$b$-tagged $t\bar t$ signal and the $Wjjjj$ background, and should
repay further study when higher statistics are achieved.

\newpage
\begin{flushleft}{\bf Acknowledgments}\end{flushleft}
This research was supported in part by the U.S.~Department of Energy
under Contract Nos.~DE-AC02-76ER00881 and DE-FG03-91ER40674, by Texas
National Research Laboratory Grant No.~RGFY93-330, and  by the
University of Wisconsin Research Committee with funds granted by the
Wisconsin Alumni Research Foundation.

\bibliographystyle{unsrt}

\newpage
\section*{Figure Captions}

\begin{enumerate}

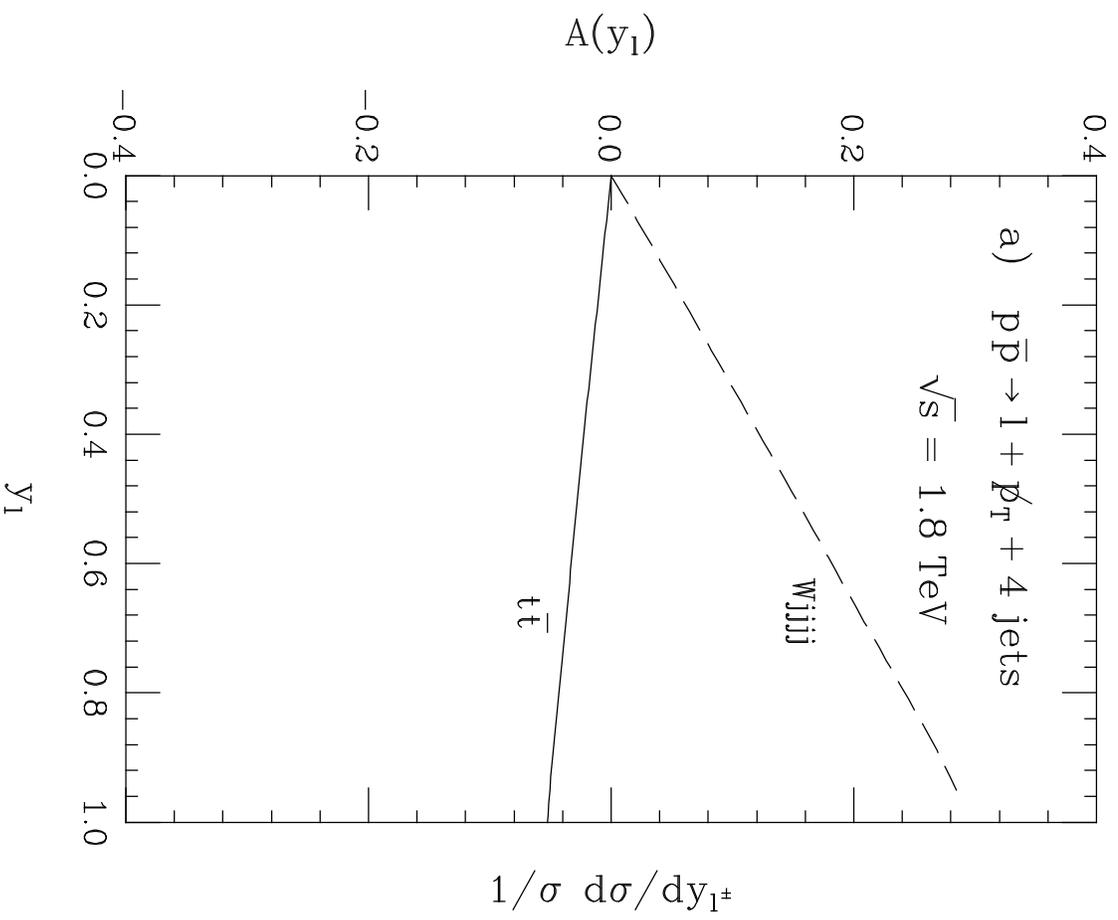
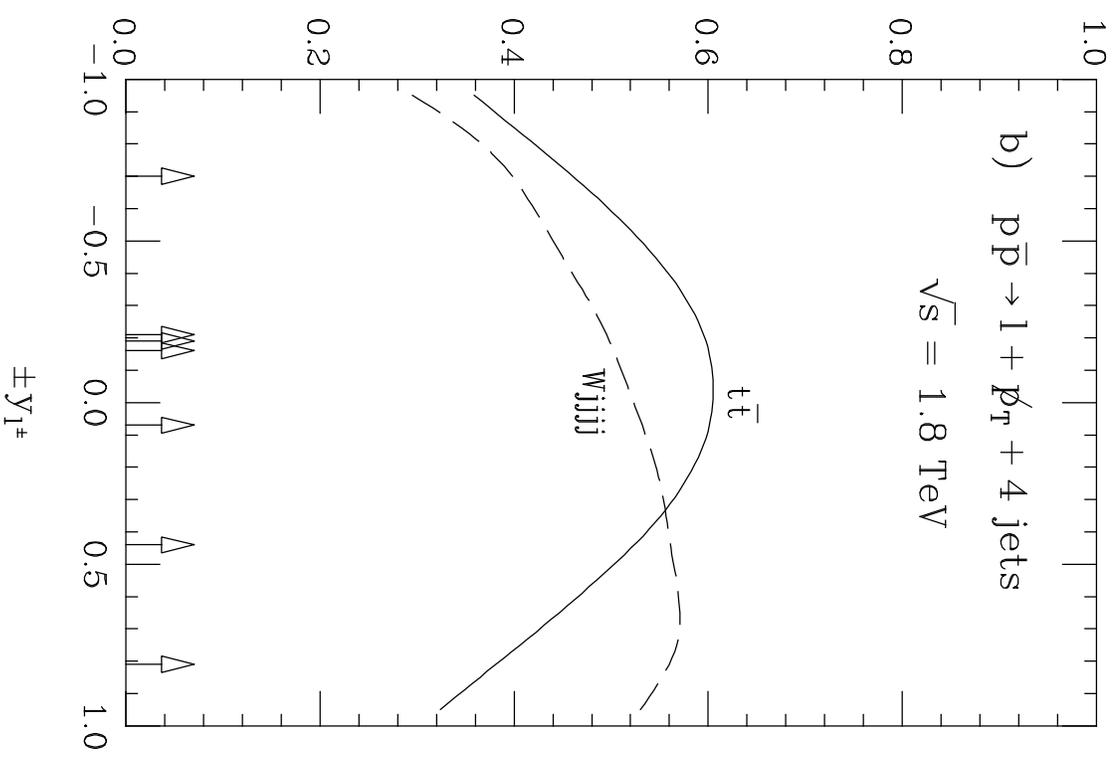
\item[Fig.~1.] (a) Forward/backward asymmetry $A(y_\ell)$ of charged
leptons versus rapidity $y_\ell$ for the $t\bar t$ signal (solid
curve)  and the $Wjjjj$ background (dashed curve) after all cuts.
(b) The corresponding lepton rapidity distributions. The $y_{\ell^\pm}$
values for the 7 reconstructed CDF events are shown by arrows.

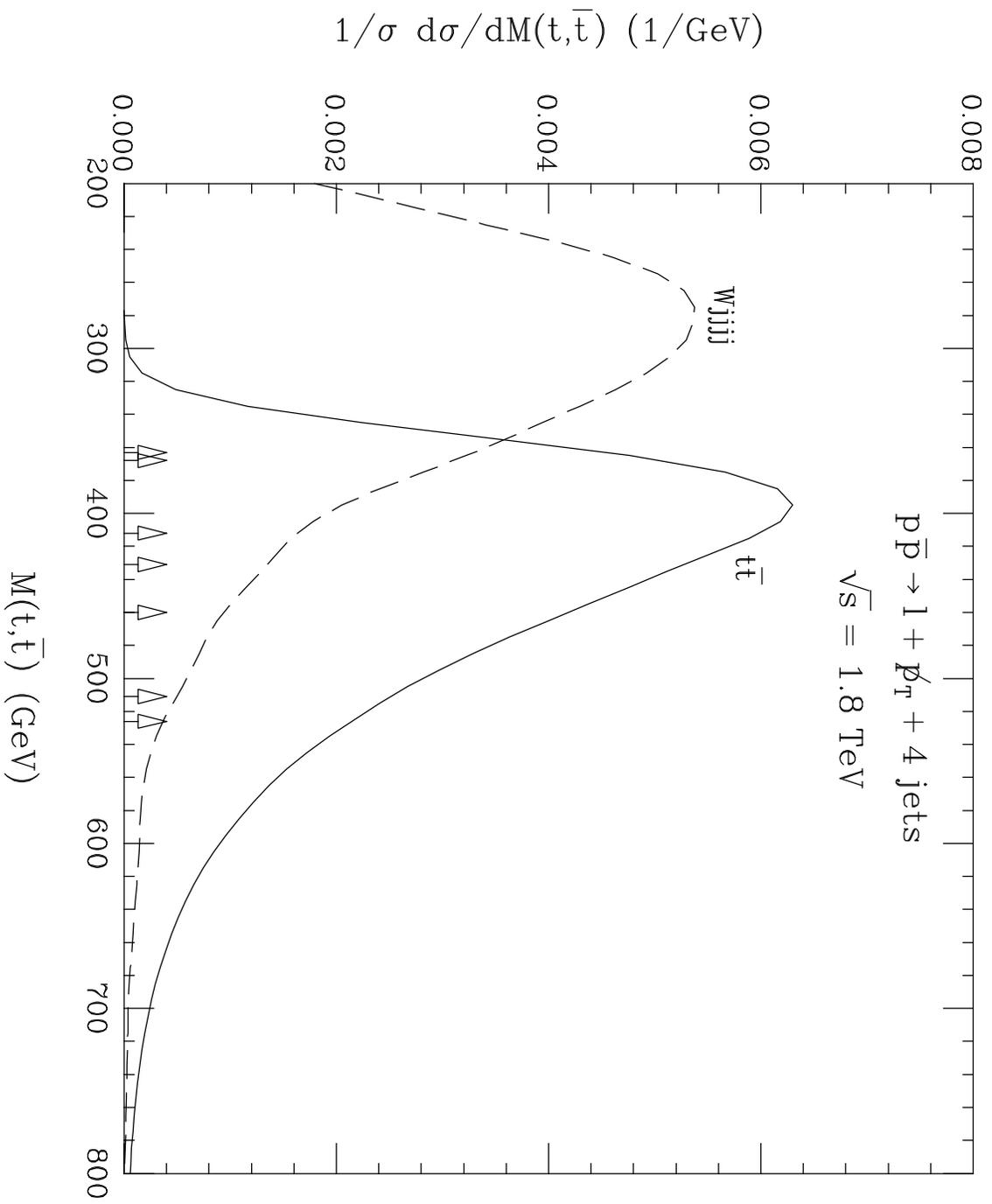
\item[Fig.~2.] Distribution of the invariant mass $m(t\bar t\,)$ for
$b$-tagged reconstructed events: the solid curve denotes the $t\bar t$
signal and the dashed curve denotes the $Wjjjj$ background. The
$m(t\bar t\,)$  values for the 7 reconstructed CDF events are shown by
arrows.

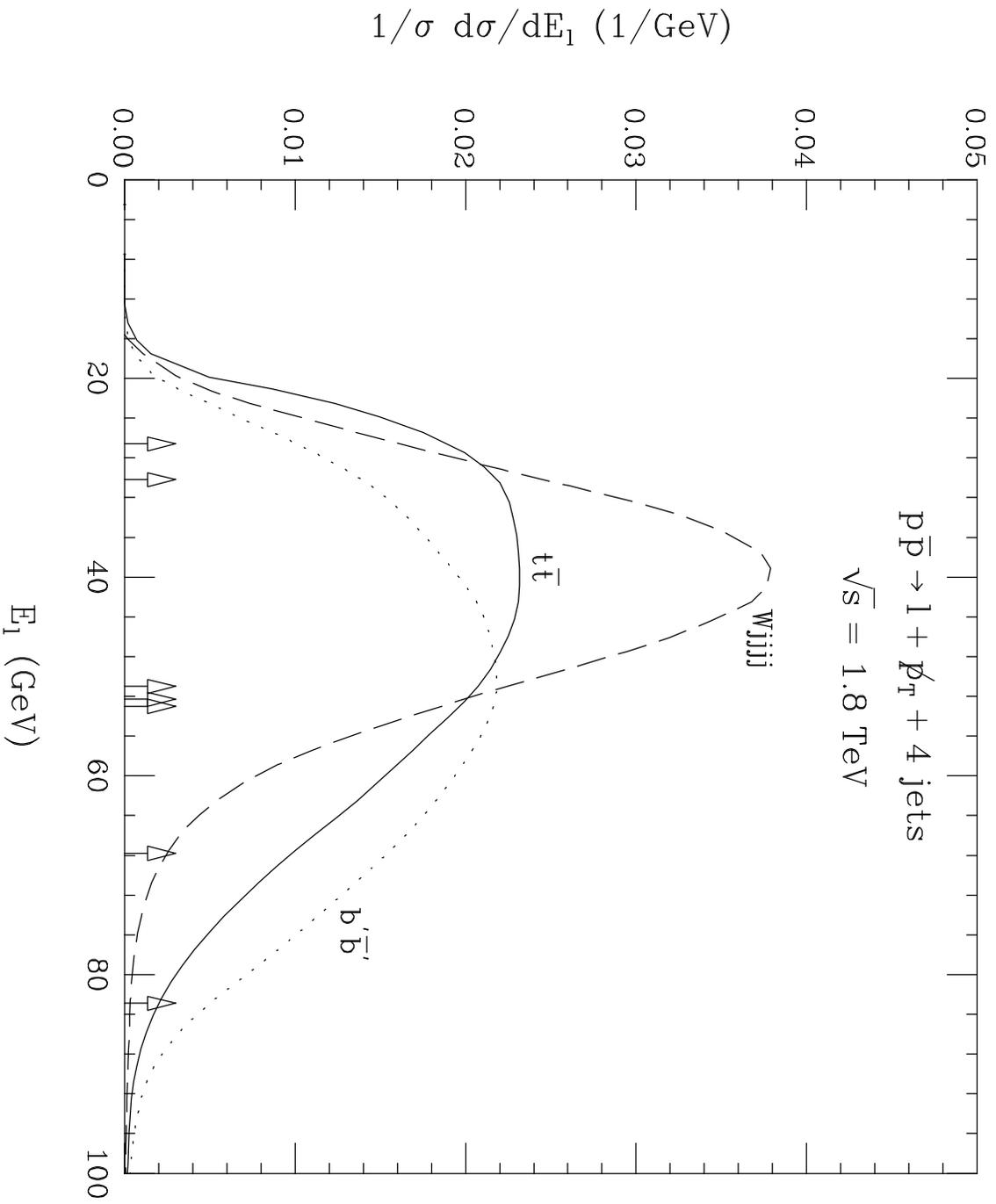
\item[Fig.~3.] Distribution of the lepton energy $E_\ell$ in the
parent quark restframe for $b$-tagged reconstructed events: the solid
curve denotes the $t\bar t$ signal, the dashed curve denotes the
$Wjjjj$ background,  and the dotted curve represents contributions
from hypothetical  $b'\bar b'$ production with $m_{b'} = m_t =
174$~GeV.  The $E_\ell$  values for the 7 reconstructed CDF events are
shown by arrows.  Note that $b' \bar b'$ events would be unlikely
to be tagged.

\end{enumerate}

\end{document}